 \documentclass{ws-ijmpa}

\usepackage[super,compress]{cite}
\usepackage{hyperref}
\usepackage{url}
\usepackage{graphicx}
\usepackage{xcolor}
\usepackage{lineno}
\usepackage{xspace} 
\usepackage{xcolor}

\usepackage[normalem]{ulem}

\definecolor{mygreen}{RGB}{15,135,30} 
\definecolor{myorange}{RGB}{180,0,220} 
\definecolor{myblue}{RGB}{0,0,200}

\definecolor{newcitecolor}{RGB}{255,0,255}

\begin{document}
\markboth{Jarosław Stasielak for the Pierre Auger Collaborration}{AugerPrime - The upgrade of the Pierre Auger Observatory}

\catchline{}{}{}{}{}
%
%%%%%%%%%%%%%%%%%%%%%%%%%%%%%%%%%%%%%%%%%%%%%%%%%%%%%%%%%%%%%%%%%%%%

\title{AugerPrime - The upgrade of the Pierre Auger Observatory}

\author{Jarosław Stasielak for the Pierre Auger Collaborration\footnote{Full author list: \url{http://www.auger.org/archive/authors_2020_09.html}} \footnote{Observatorio Pierre Auger, Av. San Mart\'{i}n Norte 304, 5613 Malargue}
\footnote{
\href{mailto:auger_spokespersons@fnal.gov}{\rm auger\_spokespersons@fnal.gov}}
}

\address{Institute of Nuclear Physics Polish Academy of Sciences\\ 
Radzikowskiego 152, 31-342 Kraków, Poland\\
jaroslaw.stasielak@ifj.edu.pl}

\maketitle

\begin{history}
\received{Day Month Year}
\revised{Day Month Year}
\end{history}

\begin{abstract}
Ultra-high-energy cosmic rays (UHECRs) are studied with giant ground-based detector systems recording extensive air showers, induced by cosmic ray particles in the atmosphere. Research at the Pierre Auger Observatory - the largest of such detectors ever built -- largely contributed to a number of breakthroughs and dramatically advanced our understanding of UHECRs. Nonetheless, the results so far are still inconclusive as neither have the sources of these most energetic particles known in the Universe been determined, nor has the origin of the unambiguously established cosmic ray flux suppression above 40 EeV been fully understood. At the same time, precise measurements of the muon component of the extensive air showers on the ground show discrepancies with the predictions of hadronic interaction models. The explanation of these puzzles, which are closely related to each other, is one of the most important goals of modern astrophysics.

The results obtained by the Pierre Auger Observatory indicate that further advances in understanding UHECRs require an improvement of the measuring capabilities of existing detectors, where the key feature is a superior separation of the muonic and electromagnetic components of air showers. AugerPrime, the ongoing upgrade of the Pierre Auger Observatory has been designed for this task. 
%As part of the upgrade, almost all of the 1661 water-Cherenkov detectors, composing the 3000 km$^2$ of the surface array of the Observatory, will be instrumented with additional scintillator and radio detector units. Moreover, new, faster and more precise acquisition electronics will be installed in all surface detectors. 
The main objective of the AugerPrime is to enhance the sensitivity of our analyses to the masses of cosmic rays, which will help to elucidate the origin of the UHECRs. In this paper we overview the main features of the AugerPrime design, its current status, and discuss the goals and potential capabilities of the upgraded Observatory.

\keywords{ultra-high energy cosmic rays; extensive air showers; AugerPrime.}
\end{abstract}

\ccode{PACS numbers: 98.70.Sa, 95.55.Vj, 95.85.Ry}

%\tableofcontents

\section{Introduction}	

The Pierre Auger Observatory\cite{ThePierreAuger}, located on a vast, high plain in Argentina, is the world’s largest 
%ever built 
detection system for the observation of
%designed primarily to study 
ultra-high-energy cosmic rays (UHECRs) (i.e. cosmic rays with energies above 1 EeV = 10$^{18}$~eV).
Detection of these most energetic particles observed in the Universe,
%UHECRs
%cosmic rays with such huge energies 
is feasible only indirectly, through observations of the so-called extensive air showers (EAS), i.e. cascades of secondary particles initiated high in the atmosphere by incoming cosmic ray particles. The air shower takes the form of a thin disk ($\sim$10 m), made up of a huge number of particles, moving with the speed of light towards the ground. The disk is slightly bent, with particles far from its central part staying a little behind. The diameter of the disk, as well as the number of particles within the cascade, change with its development in the atmosphere, and can reach several kilometers and several billions, respectively. Cascade grows, reaches a maximum and then dies out. It can be divided into three components, namely the hadronic, electromagnetic and muonic, where the last two components dominate.

The main facilities of the Pierre Auger Observatory are the Surface Detector (SD)\cite{SD}, an array of 1661 water-Cherenkov detectors (WCD) 
%arranged on a regular triangular grid of 1500 m and covering 
deployed over
%a triangular grid of 1500 m spacing over 
an area of 
3000 km$^2$, measuring the lateral distribution of particles in 
%the 
a shower at ground level, and the Fluorescence Detector (FD)\cite{FD}, with 27 telescopes, overlooking the atmosphere above the SD array.
The FD records the fluorescence light generated in the atmosphere
%the charged particles of
by the charged particles of shower through excitation of N$_2$ molecules,
%through excitation of N$_2$ molecules by the charged particles of 
%extensive air showers, 
measuring the longitudinal development of EAS 
%air shower 
in the atmosphere.
%The fluorescence light generated in the atmosphere by the charged particles of the air shower through excitation of N2 molecules is recorded by the
%the longitudinal profile of the number of shower particles (i.e. shower development in the atmosphere).
%EAS (i.e. tracking the number of shower particles at different atmospheric depths). 
%stages of 
%its
%the cascade 
%development).
The Auger Observatory is a hybrid detector utilizing
%that uses  
simultaneously 
%combines 
both of these complementary techniques to make measurements  of extensive air showers more precise.
%measure properties of extensive air showers.
%with a higher 
%enhanced 
%precision.
The FD operates only during clear, moonless nights, which limits its duty cycle to 
%less than 
$\sim$13$\%$. In contrast, the SD array 
%while the rest of  Surface Detector operates
works continuously, regardless of weather conditions.
%works 24 hours per day, 
%aside malfunction periods, reaching almost 100$\%$ duty cycle. 
Additionally, 
7 underground muon counting stations\cite{UMD,Daniel_2015} 
%the Underground Muon Detector (UMD)\cite{UMD} 
%counting muons 
and a 
several km$^2$ 
array of radio antennas\cite{AERA} 
%(Auger Engineering Radio Array, AERA)\cite{AERA}
measuring the radio signal emitted by air showers,
%several enhancements 
%extend the energy range of shower measurements down to $\sim$10$^{17}$~eV and 
provide an extra information, complementing the SD and FD data. 
%These are 
%three high elevation fluorescence telescopes (HEAT) overlooking a smaller 23.5 km$^2$ surface detector array (an infill array with 61 WCD spaced by 750 m), 
%underground muon detector (UMD) designed for muon counting and Auger Engineering Radio Array (AERA), i.e. 153 radio antennas detecting the radio signal emitted by air showers. 

\begin{figure}[t]
\centerline{\includegraphics[width=8.0cm]{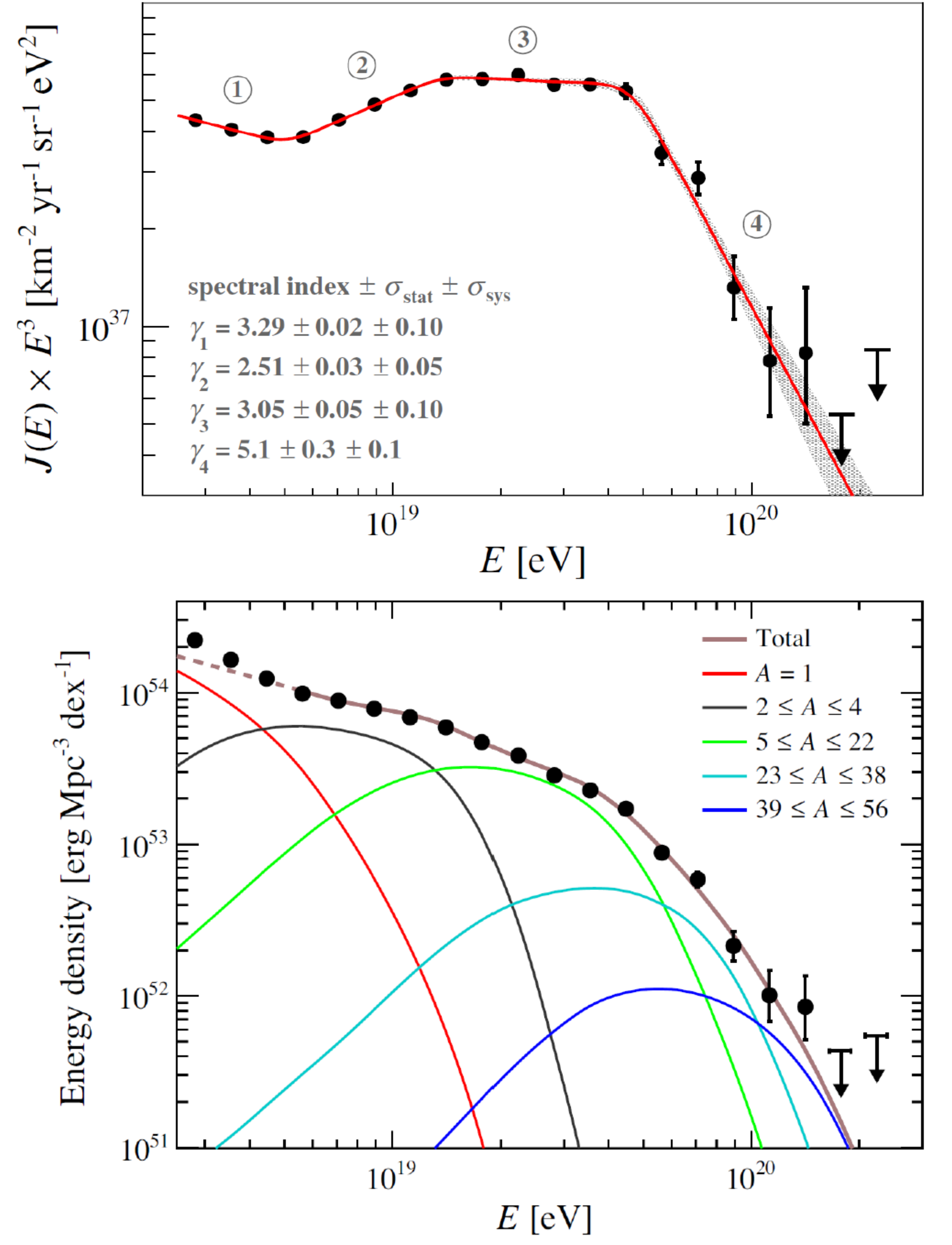}}
\caption{
Top panel: 
The Auger energy spectrum of UHECRs scaled by E$^3$. Shown are a broken power-law fit (red line) and spectral indexes of the fit for each of the energy intervals. The shaded band indicates the 90$\%$ confidence level of the fit. 
The suppression of the cosmic ray flux at $\sim$40~EeV is clearly visible;
%The shaded band indicates the statistical uncertainty of the fit. Upper limits are at the 90$\%$ confidence level. 
Bottom panel: 
Energy density of different mass components best reproducing the Auger data. The plot is obtained by simultaneously fitting
%from the best fit to 
the energy spectrum above 5 EeV and the composition-sensitive data (the distributions of both
%both 
%of 
the  atmospheric depths  of  maximum of shower development,
%of the depths of the shower maximum 
X$_{\rm{max}}$, and their fluctuations, $\sigma$(X$_{\rm{max}}$)).
%, which are mass sensitive. 
See Ref. \citen{Combinedfit} for details.
%on the UHECR sources model used. 
The mass number of nuclei 
%making up the individual components 
is denoted by A. The mass composition is dominated by heavy nuclei at the highest energies which is 
%Protons (red solid line)  
characteristic for the maximum rigidity scenario.
%acceleration efficiency 
%model. 
%Note that due to the low statistics or non-existence of the $X_{\rm{max}}$ data in and above the flux suppression region, the obtained results should be considered with reserve above $\sim$40 EeV.
%in this energy range. 
%Maximum acceleration efficiency
%parameters of the benchmark scenario used for illustration
%The dashed curve shows the energy range that is not used in the fit. 
%Only the energy range where the brown lines are solid is included in the fit. 
Figures taken from Ref.~\citen{s1}.
\label{fig1}}
\end{figure}

Research at the  Pierre  Auger  Observatory largely  contributed  to  a  number  of breakthroughs  and dramatically  advanced  our  understanding  of  UHECRs.
%
%As an example, the suppression of the cosmic ray flux above $\sim$40 EeV has been established unambiguously to high precision\cite{s1,s2,s3,s4}. Moreover, and a dipolar large-scale anisotropy in the arrival directions of cosmic rays above 8 EeV\cite{dipole,dipole2} has been observed. The magnitude and the direction of the dipole indicates an extragalactic origin of the highest-energy cosmic rays.
As an example, the suppression of the cosmic ray flux above $\sim$40 EeV has been established unambiguously\cite{s1,s2} (see Fig. \ref{fig1}, top panel). Another significant result was the observation of a dipolar large-scale anisotropy in the arrival directions of cosmic rays above 8~EeV\cite{dipole,dipole2},
%, with the magnitude and direction of the dipole 
indicating that they are indeed of extragalactic origin.
%for an extragalactic origin of the highest-energy cosmic rays.
With these advances, however, new, more specific questions have been revealed; 
%which made the results obtained so far still
%till now 
%inconclusive, 
still, the sources of
%So far
%as still
%neither have the sources of 
these most energetic particles known in the Universe 
remain to be identified and the origin of the flux suppression has to be fully understood.
%been determined, nor has the origin of the flux suppression has been  fully  understood. 
At the  same  time, precise measurements of the muon component of extensive air showers on the ground show discrepancies with the predictions of hadronic interaction models\cite{muon1,m2,m1,gora}. 
%At the  same  time, precise measurements of the muon component of air showers on the ground show more muons than is predicted in simulations\cite{muon1,m2,m1,gora}. 
%At the  same  time, precise measurements of the muon number of the extensive air showers on the ground show their 30-60$\%$ excess in relation to the predictions of hadronic interaction models\cite{muon1,m2,m1,gora}. 
%It is known as the muon deficit problem\cite{muon-deficit}. 
The explanation of these puzzles, which are closely related to each other, is one of the most important goals of modern astrophysics.

The suppression of the cosmic ray flux was predicted more than 50 years ago as a result of the particle energy losses en route to Earth due to the pion photo-production on the cosmic microwave background and is known as the Greisen-Zatsepin-Kuzmin (GZK) cutoff\cite{3,4}. In such a scenario, cosmic rays should be dominated by protons at the flux suppression region.
However, the experimental data on the cosmic ray composition collected in the Pierre Auger Observatory
%However, the experimental data on fluctuations in development of extensive air showers as well as other composition-sensitive observables
%data on the cosmic ray composition 
%collected in the Pierre Auger Observatory 
%show that the mass composition is not constant. 
show that the fraction of protons seems 
to dominate only at the energies of a few EeV, and
%mass composition becoming progressively heavier with increasing energy.
is successively replaced by heavier elements at higher energies\cite{11a,11b,11c,11d,11e,11f,11g,11l}.
This feature can be interpreted as a manifestation of the limit of the particle acceleration at the source, with the maximum energy proportional to the particle charge (it is the so-called maximum rigidity scenario). 
%(the so-called maximum rigidity scenario). 
In fact, 
the energy spectrum 
%together with 
and the composition data are better reproduced by a simple astrophysical model of the UHECR sources in the latter case\cite{Combinedfit,s1} (see Fig.~\ref{fig1}, bottom panel).
%However, interpretation of the Auger data relies heavily on the validity of the used hadronic interaction models, which have a serious problem with reproducing the number of muons in the observed showers.
%adopting a simple astrophysical model fitting the energy spectrum and the mass composition composition data suggests that the observed flux suppression is primarily an effect of the maximum
%rigidity of the sources of UHECR rather than only the effect of GZK effect\cite{Combinedfit}.
%the combined fit of the spectrum and composition data favours is better reproduced in this second scenario 
The same scenario is also supported by the lack of observation of ultra-high-energy neutrinos
%conclusion seems to be strengthened 
%by no observation of neutrinos at ultra-high energies
(which should be produced as a result of the GZK effect) -- recently more stringent upper limits for their fluxes have been set\cite{neutrino-flux}.
The GZK effect and the
maximum ridigity scenario
%models of maximum acceleration energy 
do not exclude each other, and the spectrum of cosmic rays observed may be the result of a combination of different effects. However, with the observations collected so far, we are not able to 
%discriminate between the two scenarios or estimate their 
determine which model is correct. 
%or which effect is more important. 
Composition determination of 
UHECRs
%ultra-high-energy cosmic rays 
(preferably on an event-by-event basis) at energies above the flux suppression region ($>40$ EeV) is the key to resolve this issue.

Another puzzle 
%to resolve 
is 
%a dipole anisotropy\cite{dipole2} and 
a weak correlation of cosmic rays with the distribution of nearby galaxies observed above 50 EeV\cite{dipole2,dipole,12a,12b}, 
%in the Auger data, 
which could be explained by $\sim$10$\%$ fraction of protons at these energies. 
%Small deflection of the ultra-high-energy protons in the intergalactic magnetic field should allow tracking back to their sources. 
The ultra-high-energy protons should be deflected in the intergalactic magnetic field by only several degrees, which would enable
%allow 
tracking back to their sources. 
In contrast, more massive nuclei would erase any detectable correlations.
%with UHECR
%their 
%sources. 
However, extrapolating the composition results to the highest energies, we conclude that the protons constitute much lower fraction of the cosmic ray flux than expected. 
%The dominating fraction of more massive nuclei is expected to deviate from their original directions by much larger angles, thus erasing any detectable correlation with their sources. 
%at the moment of observation. 
It is again understandable that measurement of cosmic ray composition at energies above 50 EeV is crucial for further progress in this investigation. The ability to identify primary protons in the observed 
UHECRs
%ultra-high-energy cosmic rays, 
would provide
%, thanks to their relatively small magnetic deflections, 
a way to reliably point to the sources, and thus possibly establish the proton astronomy.

Determination of cosmic ray composition
is closely related to air shower physics and thus
%depend on the
relies heavily on the 
%properties of 
hadronic interaction models.
%, so that any inaccuracies in these models may influence the interpretation of the experimental results.  
Any inaccuracies related to extrapolation of their properties, 
%of hadronic interaction models,
from energies at which they are measured at terrestrial particle accelerators, to the ultra-high energies may influence the interpretation of the 
mass-sensitive data.
%experimental results.
%Precise  theoretical  predictions  are  crucial  for correct interpretations of the mass composition 
%observational 
%data. 
%Unfortunately, properties of hadronic interactions measured at terrestrial particle accelerators have to be extrapolated to the ultra-high-energies, contributing to systematic uncertainties of the final results.
%the energy range relevant to UHECR
%ultra-high-energy cosmic ray 
%studies is far beyond the energies accessible at terrestrial particle accelerators, at which the properties of hadronic interactions have been measured. Thus, an extrapolation of hadronic interaction properties to higher energies is necessary, contributing to systematic uncertainties of the final results.
%Extensive air shower simulations have to be made and compared to the experimental distributions. Clearly, the results depend critically on the accuracy of these simulations, which in turn depend on the assumptions made concerning interaction cross sections, inelasticity of interactions, multiplicities and distributions of produced particles, etc. 
The 
%observed 
number of muons seen experimentally in air showers is 30-60$\%$ larger than the numbers
%predicted by hadronic interaction models\cite{muon1,m2,m1,gora}.
obtained in the simulations\cite{muon1,m2,m1,gora}.
%The measurements of showers made so far indicate strongly that the number of muons seen experimentally in air showers is 30-60$\%$ larger than predicted in the current hadronic interaction models\cite{muon1,moun2} [9]. 
%None of the models used
%used in the simulations 
%can reproduce the observed particle distributions. 
%This leads us 
This indicates a poor understanding of hadronic interaction models at the highest energies and shows
that
%our understanding of hadronic interactions at the highest energies is far from being satisfactory.
further advances 
%in the 
%understanding
%field of UHECRs 
require an improvement of the measuring capabilities of existing detectors, where the key feature
%to resolve this problem.
%This bring us 
%once more
%again 
%to the conclusion that we need
is a superior separation of the muonic and electromagnetic components.
%to study the muonic and electromagnetic components separately  
%to resolve this problem, which 
This will simultaneously help us with the mass composition study.
%interpretation of the mass composition data. 

%the key feature is a superior separation of the muonic and electromagnetic components of air showers.

The ongoing upgrade of the Pierre Auger Observatory, called AugerPrime\cite{prime} (Primary cosmic Ray Identification
with Muons and Electrons), has been designed for this task. 
%It will improve the mass composition sensitivity of the SD on a shower-to-shower basis, to explore the cosmic ray composition at energies above $10^{19}$ eV
It will enhance our ability to measure the mass composition of cosmic rays above $\sim10^{19.5}$ eV (possibly on event-by-event basis), providing information for more accurate analysis of air shower development and allowing:
%It will allow us: 
\begin{itemize}
    \item to elucidate the mass composition and the origin of the flux suppression at the highest energies, i.e. the differentiation between the GZK and maximum rigidity scenarios.
    %of energy loss effects due to cosmic ray propagation and the scenario of maximum rigidity.
    %energy of particles injected by astrophysical sources.
    \item  to reach a sensitivity to a contribution of protons as small as 10$\%$ in the flux suppression region, thus evaluating the possible existence of a small fraction of protons at these energies and assessing the feasibility of charged particle astronomy.  
    \item to improve 
    %not only our understanding of the properties of cosmic rays, but also the
    our knowledge about hadronic interactions: increasing the accuracy of existing models extrapolated to the extreme observed energies.
\end{itemize}

As a part of the upgrade, the SD water-Cherenkov tanks
%, composing the 3000 km$^2$ of the surface array of the Observatory, 
will be equipped with additional scintillator (SSD) and radio detector (RD) units. In addition, new, faster and more precise acquisition electronics along with a small PMT will be installed in all surface detectors.
%As a part of the Observatory upgrade, 
It is also planned to relax the restrictions on the FD operations, which will increase its duty cycle.
Finally, the AugerPrime will be supplemented with the upgraded Underground Muon Detector (UMD).
%small array of underground muon detectors (UMD).

\section{New Surface Scintillator Detector (SSD)}

To use the full statistics of the SD data to analyze the composition of cosmic rays, the Pierre Auger Collaboration has made the decision to upgrade the SD by placing surface scintillator detectors on top of %$\sim$1500 (
almost all (beside stations located at the outskirt of the array) of the existing SD water-Cherenkov detectors\cite{prime}.
%Muons have larger energy deposits in water than electromagnetic particles, and at the same time both components deposit on average the same amount of energy in the scintillator. In this sense, the WCD is more sensitive to muons, whereas the SSD to the electromagnetic component of shower.
The plastic scintillators and water-Cherenkov detectors responses to shower components are different.
Muons have larger energy deposits in water than electromagnetic particles, and at the same time both of these components deposit on average the same amount of energy in the scintillator. In this sense, the WCD is more sensitive to muons, whereas the SSD to the electromagnetic component of shower.
Given the different sensitivities of 
%WCD and WCD
%plastic 
scintillators and water-Cherenkov detectors 
to electrons, photons, and muons of extensive air shower, the combination of measurements from these two detector types will provide muon content information (on event-by-event basis), which is vital for cosmic ray mass composition studies and improved energy determination. 
The disentangling electromagnetic and muonic components at ground is the best available method for the mass composition measurements, when the direct optical observation of the depth of shower maximum, X$_{\rm{max}}$, by the FD is not possible 
(which is the most reliable method of mass composition determination, but very limited at the highest energies due to the small FD duty cycle).
%(X$_{\rm{max}}$ measurement by FD is the most reliable method of mass determination available so far).
Comparing the signals from the two detectors
will allow us to separate the contributions from muons and electromagnetic particles with a precision
unavailable before. 
The Observatory upgrade will allow measurements of the composition of cosmic rays at energies above $\sim$40~EeV, i.e. above the flux suppression region, which have been unattainable until now.

\begin{figure}[t]
\centerline{\includegraphics[width=12.9cm]{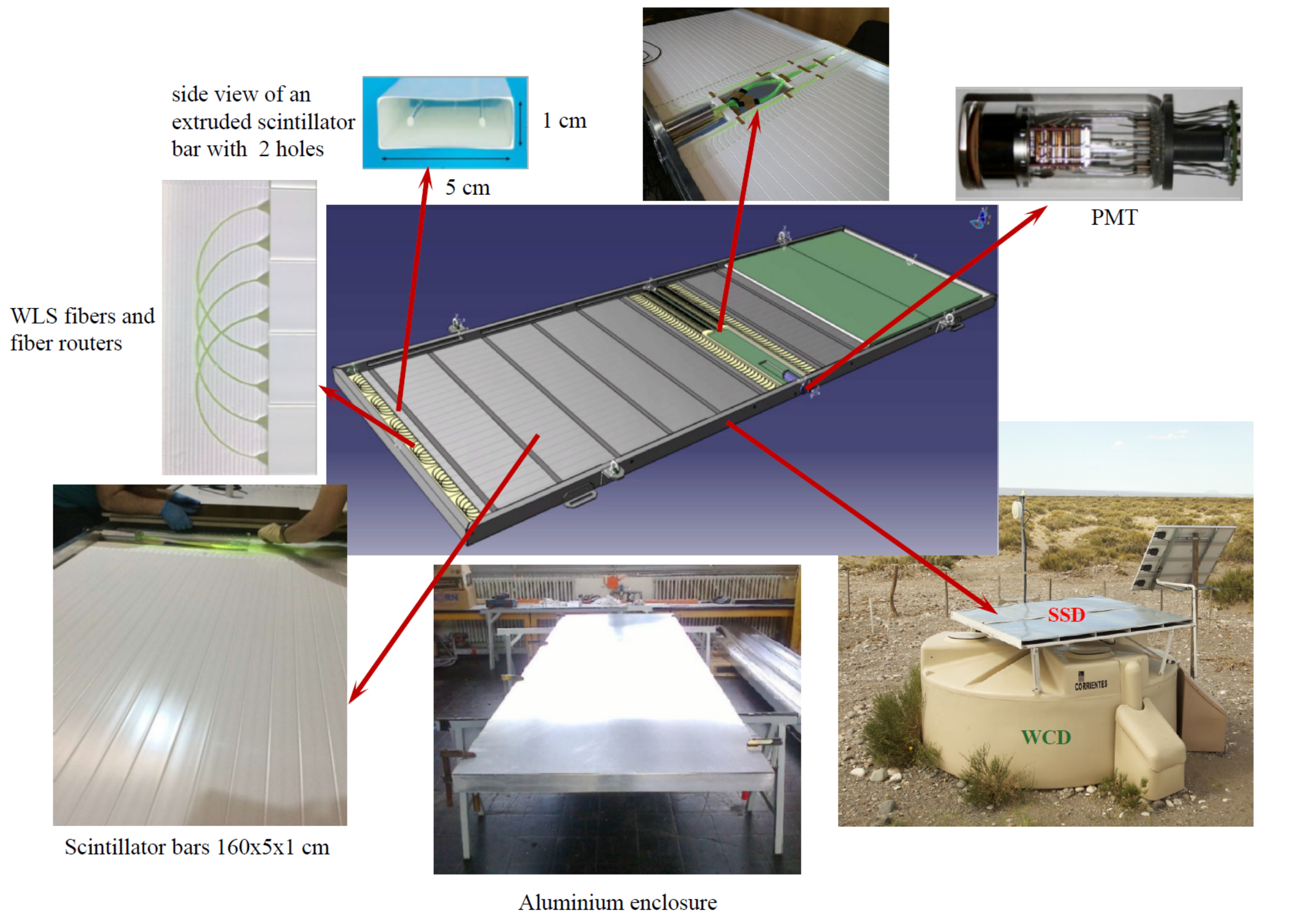}}
\caption{Layout of the surface scintillator detector (SSD) module. See description in the text. \label{SSD-constr}}
\end{figure}

A complementary measurement of shower particles is provided by a plastic scintillator plane above the existing water-Cherenkov detectors. 
The design of the SSD is simple, reliable and the SSD modules can be easily deployed over the full 3000~km$^2$ area of the Surface Detector. 
The SSD module is composed of a light-tight, waterproof, aluminium enclosure of 3.8 m x 1.3 m, housing 48 scintillator bars made of extruded polystyrene, which are the active part of the detector. They are placed symmetrically on
both sides of the detector. Each scintillator bar of dimension 160~cm x 5 cm x 1 cm, with an outer reflective layer of TiO$_{\text{2}}$,  has two co-extruded channels through which wavelength-shifting optical fibers are routed.
The fibers are laid
%disposed 
in a U-shape configuration that maximizes light yield and uniformity. All ends of the fibers, collecting the scintillation light produced by
the shower particles, are bundled towards a single photomultiplier tube (PMT) located in the central part of the module. The design of the SSD was tested and validated in the Pierre Auger Observatory.
The layout of an SSD module  and its placement above a WCD detector is shown in Fig.~\ref{SSD-constr}.

\begin{figure}[t]
\centerline{\includegraphics[width=12.9cm]{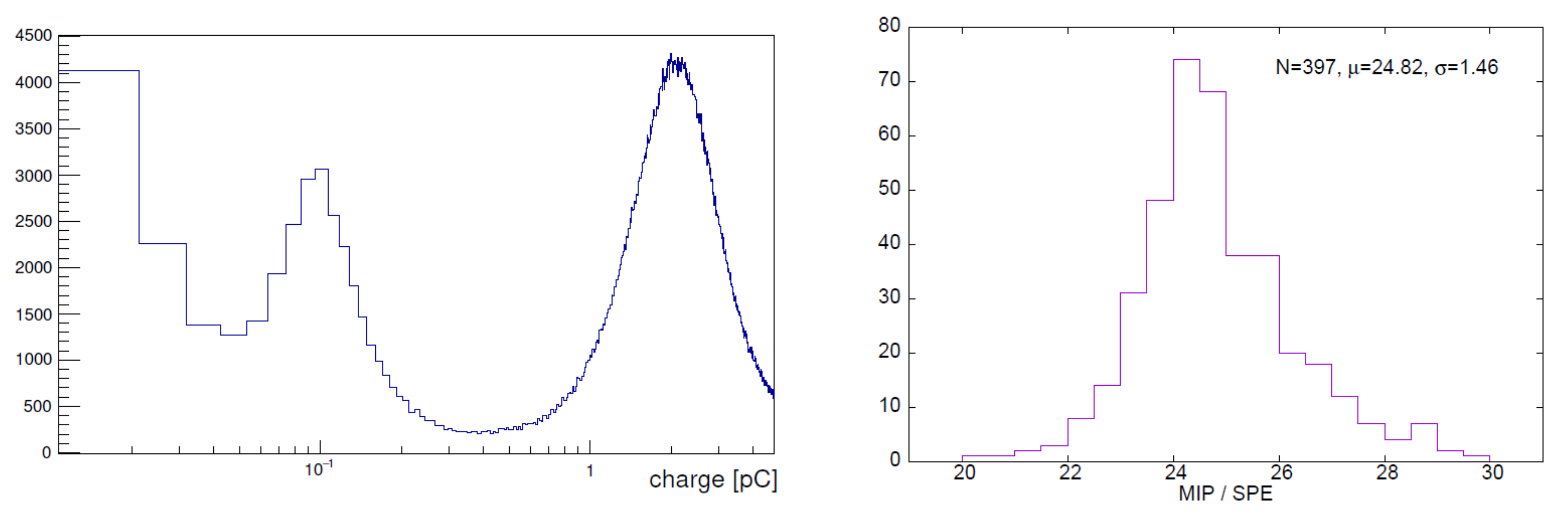}}
\caption{Left panel: Histogram of signals recorded in a test measurement of one SSD module. The peaks corresponding to electronics noise, SPEs and MIPs are shown from left to right, respectively; Right panel: Distribution of the MIP to SPE ratio for SSDs made in one of the assembly sites.
Figures taken from Ref. \citen{pekala}.
\label{MIP}}
\end{figure}

The assembly of the SSD modules has been distributed among several institutions of the Pierre Auger Collaboration. The large-scale production started in 2017, and it is now completed.
All SSDs
%After production, they 
were tested at each construction site upon completion, exploiting  atmospheric muon tomography to measure the minimum ionizing particle (MIP) as well as 
to check the SSD modules uniformity\cite{pekala}.
The test measurements allow identification of both MIPs and single photoelectrons (SPEs) signals (see Fig.~\ref{MIP}, left panel). The mean charge 
of the MIP and SPE 
is determined by fitting peaks to the histograms obtained during the tests. The ratio of these values reflects the overall efficiency of SSD (efficiency of the key detector components at generating, collecting and transmitting the light).
An example distribution of the MIP to SPE ratio obtained from one institution is shown in the right panel of Fig.~\ref{MIP}.

%On the other hand, 
The shape of the MIPs peak (Fig.~\ref{MIP}, left panel) tests the uniformity of SSD response over its active area. A deformation of this peak would indicate that a large part of the detector is deficient.
More detailed tests of selected SSD modules, based on a different method, showed that signals from individual scintillator bars deviate by no more than 10$\%$ from the average signal.

%Signals from opposing ends of a scintillator bar differ by ∼5%, as follows from the light attenuation in the fiber.

%the uniformity of the measured signals is 5$\%$ along the bars and 10$\%$ between bars

%Signals from the WCDs are expressed in units of VEM (vertical-equivalent muon), which corresponds to the average signal resulting from a vertical and centered muon traversing the tank, measured as the integrated PMT pulse over time, thus a “charge”. The term MIP (minimum ionizing particle) is used to express the signals from the SSD and it refers to the average charge deposited by a MIP.
 
As a part of the Auger upgrade,  new, faster and more precise acquisition electronics will be installed in all SD stations, thus replacing the so-called Unified Board (UB) with an Upgraded one
%Unified Board 
(UUB)\cite{nitz}. The UUB will enable to simultaneously process signals
%the SD stations will be upgraded with new electronics, which will process all signals
from the water-Cherenkov detectors, surface scintillator detectors, radio detectors, and underground moun detectors and will increase the data quality by providing:
\begin{itemize}
\item faster sampling of ADC traces 
%at a rate of 120 MHz compared to the current
%rate of 40 MHz
%(120 MHz instead of 40 MHz) 
(40 MHz $\rightarrow$ 120 MHz) 
and more precise absolute timing accuracy from new GPS receivers 
%(4 ns instead of 12 ns),
(12 ns $\rightarrow$ 4 ns), 
which is better suited for counting muons (distinguishing sharp, large muon pulses close in time)
\item faster data processing and more sophisticated local trigger (more powerful processor and FPGA)
\item larger dynamic range (10 bits $\rightarrow$ 12 bits)
\item more FADC channels (6 $\rightarrow$ 10)
\item improvement in the calibration and monitoring capabilities
\item backwards-compatibility with the old design (similar power consumption, hardware interfaces, etc.).
\end{itemize}

\begin{figure}[t]
\centerline{\includegraphics[width=12.8cm]{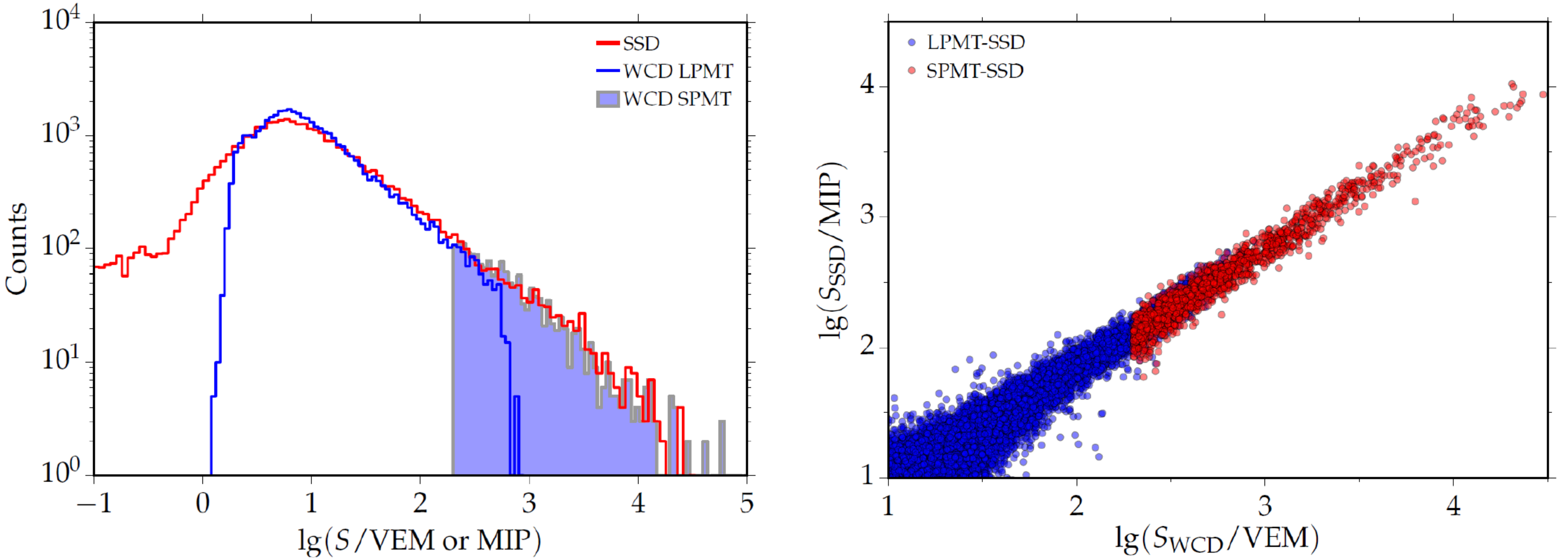}}
\caption{Left panel: Distributions of charge measured in one of the upgraded SD stations by SSD (red) and WCD (blue) signals. The filled histogram shows
signals from the small PMT; 
Right panel: Correlation between SSD and WCD signals (measured in MIP and VEM respectively) using the large PMTs (blue) and the small PMT (red). \label{SSDa}}
\end{figure}

Moreover, each WCD will be quipped with an additional 
%4th 
small photomultiplier (SPMT), with a low-gain, and an active area of about $1\%$ of the currently working large PMT (LPMT). The SPMT will register large pulses from very close showers that saturate the LPMT signal,
 %of the currently working LPMT, 
 enlarging
%enlarge 
the dynamic range of  the detector
%measurable signal intensities 
by a factor of 32 (up to $\sim$20~000 VEM, the energy equal of Vertical Equivalent Muon). This is demonstrated in Fig.~\ref{SSDa}, which shows distributions of charge measured in one of the upgraded SD stations (left panel)
and the correlation between the signals in one of the WCD and in the corresponding SSD
%SSD signal and signals measured by the WCD's large and small PMTs 
(right panel). Signals are expressed in
physical units~-- VEM for the WCD and MIP for the SSD.
The dynamic range in the upgraded SD is nicely covered by the LPMTs up to the saturation and then extended to the highest SSD signals by the SPMT. 
Larger dynamic range will allow for determination of the particle densities closer to the shower core, e.g. down to 250 m for a 10 EeV shower. 
%Both the SSDs and the electronics upgrade can be easily deployed, and will have only minimal impact on the continuous data taking of the Observatory

%AugerPrime Engineering Array (12 stations) and Pre-production SSD array (77 stations) %are already operational.

An Engineering Array consisting of the 12 upgraded SD stations has been collecting data since 2016. In addition, a pre-production SSD array of 77 stations is fully operational
%have been collecting data 
since March 2019. The upgraded stations operate with good stability. The results obtained by these arrays are very encouraging. They
%obtained by these arrays so far
%The data collected so far 
demonstrate the quality of the new detectors and the physics potential of the upgraded SD array.
More than 1000 SSD modules (out of $\sim$1500) have been already deployed in the field. Further deployment is in progress and will be finished soon.

\section{The Radio Upgrade}

%\begin{figure}[t]
%\centerline{\includegraphics[width=8.0cm]{plots/radio-foot.pdf}}
%\caption{Radio emission footprints in the 30-80 MHz frequency band as simulated with CoREAS\cite{Huege:2013vt}. While the radio emission footprint is small for air showers with zenith angles up to $\approx 60^{\circ}$, it becomes very large for inclined showers with zenith angles of 70$^o$ or larger. The white rectangle characterizes the size of the 50$^o$ inset. Figure taken from Ref. \citen{radio-foot}. \label{radio-foot}}
%\end{figure}
 
\begin{figure}[t]
\centerline{\includegraphics[width=4.5cm]{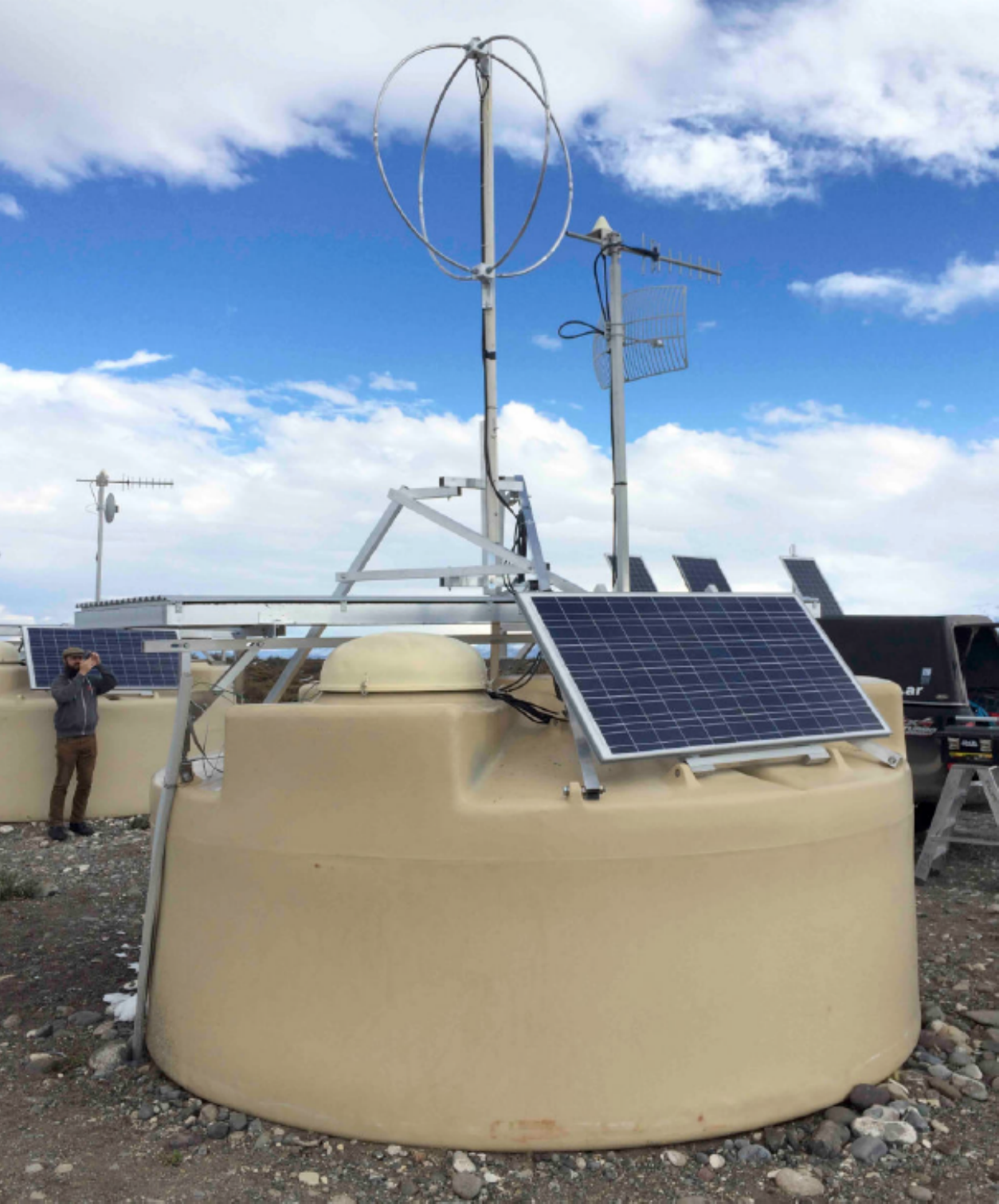}
\includegraphics[width=8.0cm]{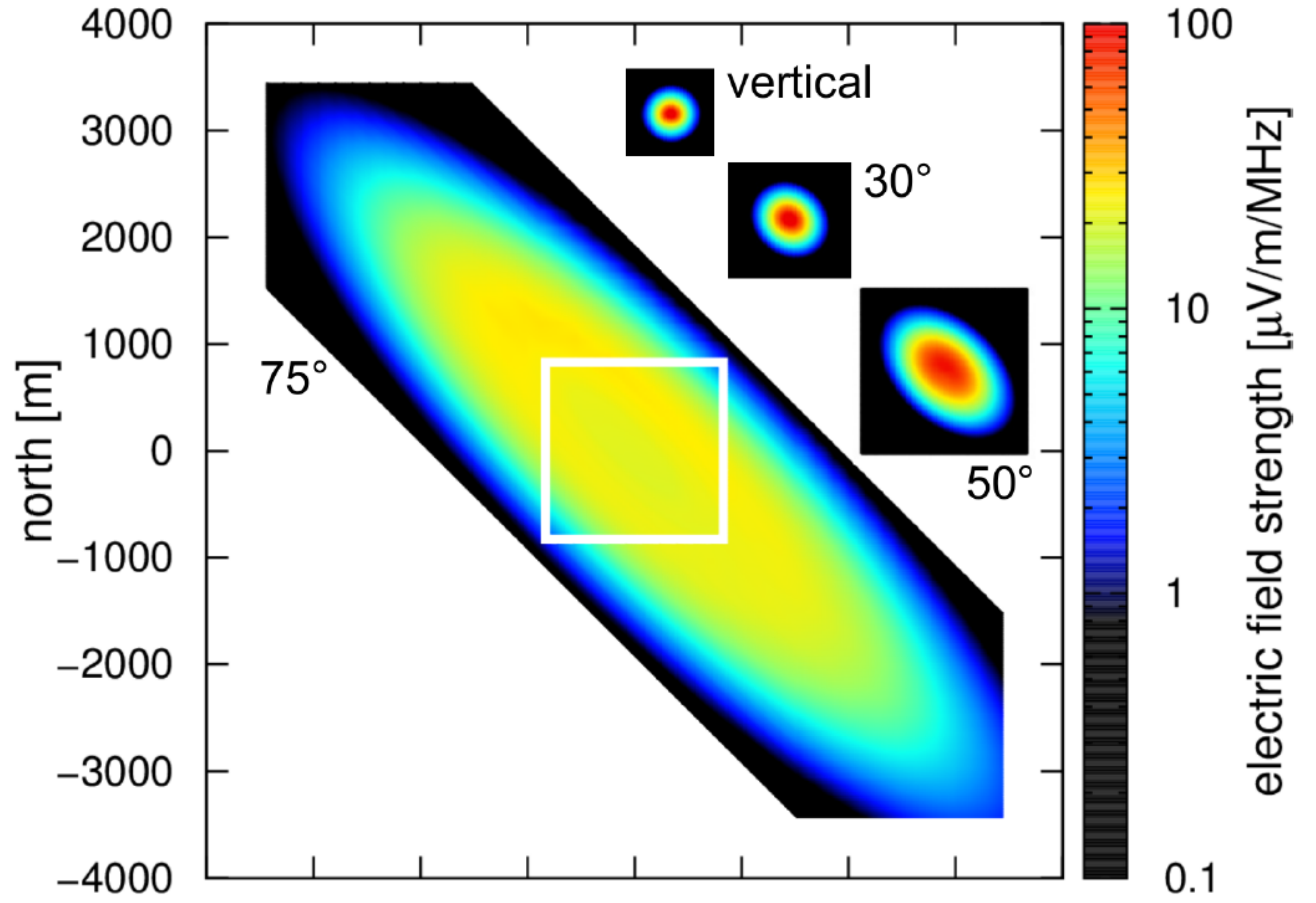}}
\caption{Left panel: An upgraded SD station, including the new scintillator detector and new radio antenna on top of the Water-Cherenkov detector; Right panel: Footprint of the shower radio emission on the ground
%Radio emission footprints 
in the 30-80 MHz frequency band as simulated with CoREAS\cite{Huege:2013vt}. While the radio 
emission 
footprint is small for air showers with zenith angles up to $\approx 60^{\circ}$, it becomes very large for inclined showers with zenith angles of 70$^{\circ}$ or larger. The white rectangle characterizes the size of the 50$^{\circ}$
inset. Figure taken from Ref. \citen{radio-foot}.  \label{WCD+SSD+RD}}
\end{figure}

Extensive air showers consist of charged particles that undergo acceleration in the atmosphere and therefore they are %known to be 
a source of radio emission in the frequency range of tens MHz up to several GHz\cite{Schr_der_2017}. The radio emission mainly arises due to 
%is dominated by 
the geomagnetic 
%effect, i.e. 
deflection of the electrons and positrons in the shower, however, 
%the radiation arising from time-varying transverse currents induced by
%in the geomagnetic field, with an admixture of 
the Askaryan effect related to the time variation of the net charge excess in the shower front, is also important.
For several years now, the Auger Engineering Radio Array (AERA)\cite{refId0}, an array of more than 150 radio detector stations covering an area of about 17 km$^2$, 
%co-located with the SD Infill array, 
has been successfully recording the radio emission of air showers.
To fully utilise detection capabilities of 
%air showers by 
the radio technique, it was decided to extend the existing Radio Detector to the entire SD array. Thus, 
%operated at the Auger observatory.
%It was with this detector that the maturity of the radio technique was demonstrated for the first time
%by measuring air showers 
%energy 
%with the Radio Detector (RD) alone, without external triggering\cite{PhysRevLett.116.241101}.
%,PhysRevD.90.122005}.
%triggering from an external detector\cite{PhysRevLett.116.241101,PhysRevD.90.122005}.
%in addition to the SSD,  the
each water-Cherenkov detector of the SD array will be equipped both with an SSD and a radio antenna, 
%mounted on its top surface, 
forming the largest radio array in the world, covering 3000 km$^2$.
An already upgraded SD station, with a scintillator detector and a radio antenna installed is shown in Fig.~\ref{WCD+SSD+RD} (left panel). 

The radio array will be composed of 
%consists of 
the short aperiodic loaded loop antennas (SALLAs) detecting the radio emission from air showers in the frequency range of 30 to 80 MHz.
The SALLA has been developed to provide a minimal design that fulfills the need for both ultra-wide band sensitivity and low costs for production and maintenance of the antenna in a large-scale radio detector. The compact structure of the SALLA makes the antenna robust and easy to manufacture. 
%Its sensitivity has a flat distribution in frequency.
%has a flat distribution in frequency.
Its sensitivity is a flat function of radio frequency.
%Its structure  creates a sensitivity, which is flat as a function of frequency. 
The antenna measures along two polarization directions, oriented orthogonal to each other. 

The new Radio Detector will operate together with the upgraded SD, forming a unique setup to measure the properties of 
air showers.
%cosmic rays.  
The size of the footprint of radio emission on the ground 
%is quite small for vertical showers (several 100 m) and 
increases with the shower zenith angle (see Fig.~\ref{WCD+SSD+RD}, right panel), 
%reaching
covering 
areas of the order of 100 km$^2$ for 
%almost horizontal 
very inclined 
showers. 
Due to the size of the radio footprint, observations of air showers by a sparse antenna array with 1.5 km spacing is only feasible for inclined showers, i.e. those with zenith angles larger than $\sim60^{\circ}$.
The  RD  will  be  useful  for studying  inclined  showers, for  which the SSD efficiency drops due to geometrical reasons.
Supplementing the water-Cherenkov stations 
with both SSD and RD detectors will extend the shower measurement capabilities over the full angular range.
The combination of WCD and RD will be used to measure the ratio of the electromagnetic energy content and the number of muons (providing separation of the electromagnetic and muonic components) for inclined air showers at the highest energies. It is similar to what has previously been done with the combination of WCD and Fluorescence Detector, however, 
with  much higher event statistics due to the 100$\%$ duty cycle of the Radio Detector.
%It will improve the science capabilities of the Auger Observatory, by providing electron-muon separation for inclined air showers at the highest energies. 

The Auger Radio Detector is a natural next step towards future cosmic-ray experiments.
%, applying hybrid detection techniques by combining radio antennas with e.g. segmented water-Cherenkov detectors. 
%The Auger Radio Detector 
It will allow to evaluate the detector technology, establish reconstruction methods, and study the physics performance of huge sparse radio arrays.

%The upgrade includes the installation of radio antennas on each of the 1661 water-Cherenkov detectors of the array. The
%main objective of the radio upgrade (Radio Detector) is to measure horizontal air showers and to determine the properties of cosmic rays up to the highest energies. 
%The combination of water-Cherenkov detectors and radio antennas will provide muon-electron separation for horizontal air
%showers at the highest energies.

\section{Upgrade of the Underground Muon Detector}

%The underground detectors, that are sufficiently shielded against the electromagnetic component, provide the most direct measurements of muon content of air showers.

\begin{figure}[t]
\centerline{\includegraphics[width=12.0cm]{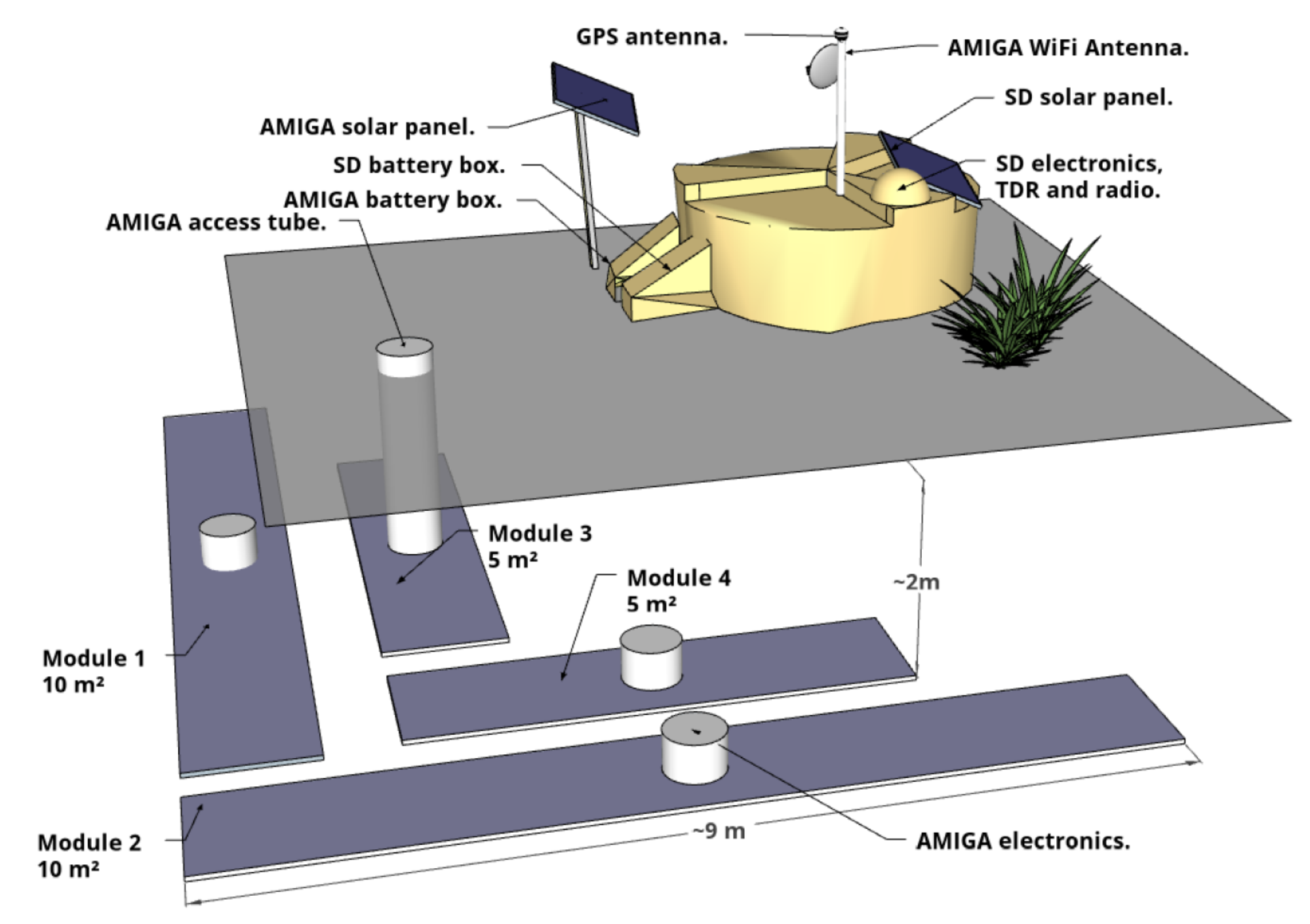}}
\caption{Underground muon detector general overview. Both surface and underground detector, are shown in their arrangement during the prototype phase. In the final design, the 30 m$^2$ detector is splitted into three 10 m$^2$ modules.
 \label{amiga2}}
\end{figure}

The most direct measurement of the muon content of air showers
%the muon 
%number 
%measurements relies on the usual counting of muons by an
%is based on the muon counting by
%counting the muons by 
is provided by the underground detectors, buried under a several-meter layer
of soil,
%buried under a few-meter layer
%of soil 
that are 
%that are 
sufficiently 
shielded against the electromagnetic component. 
It would be impractical, however, to install 
%them
such detector
%dig holes for an underground detector 
under all of the 1661 SD stations,
%water-Cherenkov tank, 
that are distributed over a large area, with difficult access in some parts. 
Therefore, it was decided to build a smaller 
%underground detector 
array by
extending the existing Underground Muon Detector, currently consisting of 7 AMIGA\footnote{Auger Muons and Infill for the Ground Array} muon
stations\cite{Daniel_2015,UMD}, to an area of the 
%existing 
%23.5 km$^2$
%of the existing 
SD Infill, i.e. a small WCD array designed to extend the energy range of shower measurements of the Auger Observatory down to $\sim10^{17}$~eV. 

\begin{figure}[t]
\centerline{\includegraphics[width=12.8cm]{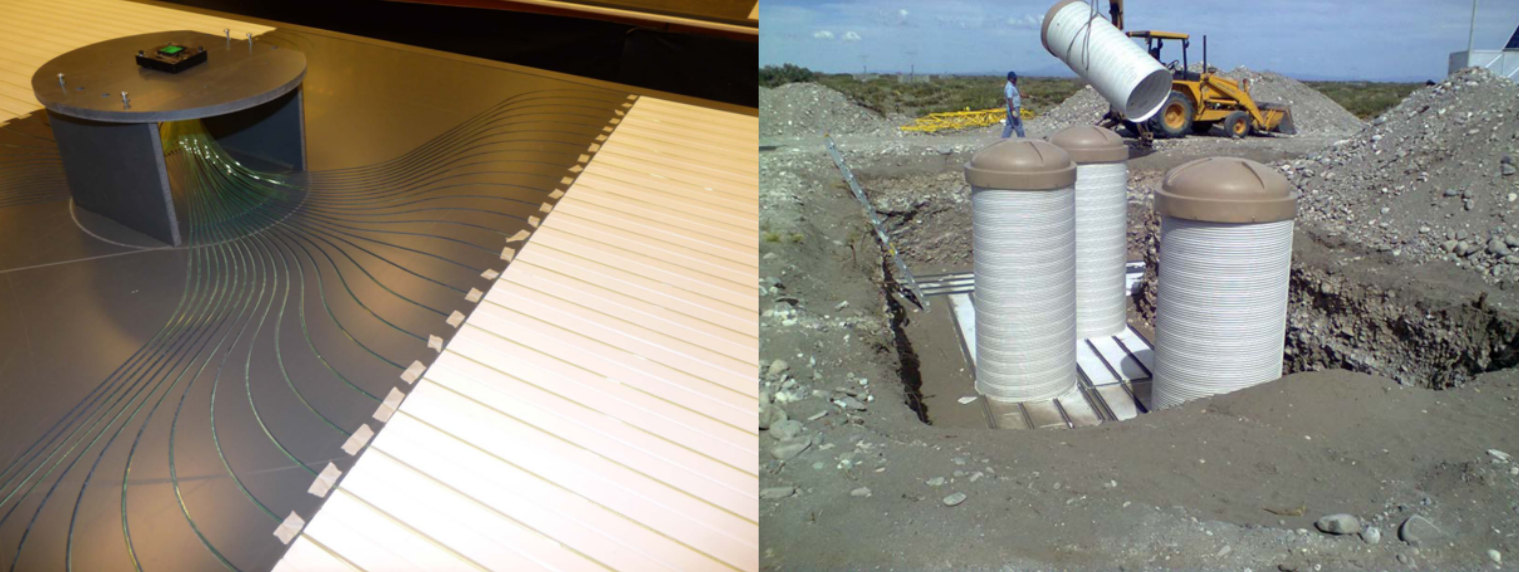}}
\caption{Left panel: Scintillator module. The black
optical connector concentrates the 64 optical fibers coming from the
32 scintillator bars at each side of the detector; Right panel: Deployment of an AMIGA muon detector. The shown service-tubes are to provide access to the electronics for development and
maintenance purposes. 
 \label{amiga}}
\end{figure}

Each 
%\textcolor{red}{\sout{water-Cherenkov tank}
SD station of the Infill will be supplemented with the AMIGA muon detector (UMD station), consisting of 3 scintillator modules of 10 m$^2$ area, buried at a depth of 2.3 m, 
%beside to it, 
at a distance large enough from the water tank to avoid shadowing 
%from the water tank, 
(to guarantee uniform shielding), see Fig.~\ref{amiga2}.
%of the muon detectors.
The 
%10 m$^2$ 
scintillator module
%of the UMD stations 
is made of 64 scintillator strips (4 m long, 4.1 cm wide, 1 cm thick).
%Scintillation 
Light collected at each strip is guided, using wavelength shifting fibers, towards a 64 channel silicon photo sensor\cite{Aab_2017} located in the middle of the module (see Fig.~\ref{amiga}, left panel). 
The segmented structure of the scintillator module allows for direct counting of muons.
The 
%entire 
Underground Muon Detector will consist of 61 stations deployed 
%beside the water-Cherenkov tanks of the SD Infill 
on a 750 m triangular grid of the SD Infill, instrumenting its entire area of 23.5 km$^2$. Deployment of an UMD detector is shown in Fig.~\ref{amiga} (right panel).

Comparing the measurements obtained by the surface detectors with those from a small array of underground muon detectors should significantly improve the accuracy of the AugerPrime results. 
The UMD will provide important direct muon measurements of a sub-sample of showers, along with their  time structure. It will be used for verification and fine-tuning of the methods used to extract muon information from the SSD and WCD measurements.

\section{Increase of the FD up-time}

\begin{figure}[t]
\centerline{\includegraphics[width=12.8cm]{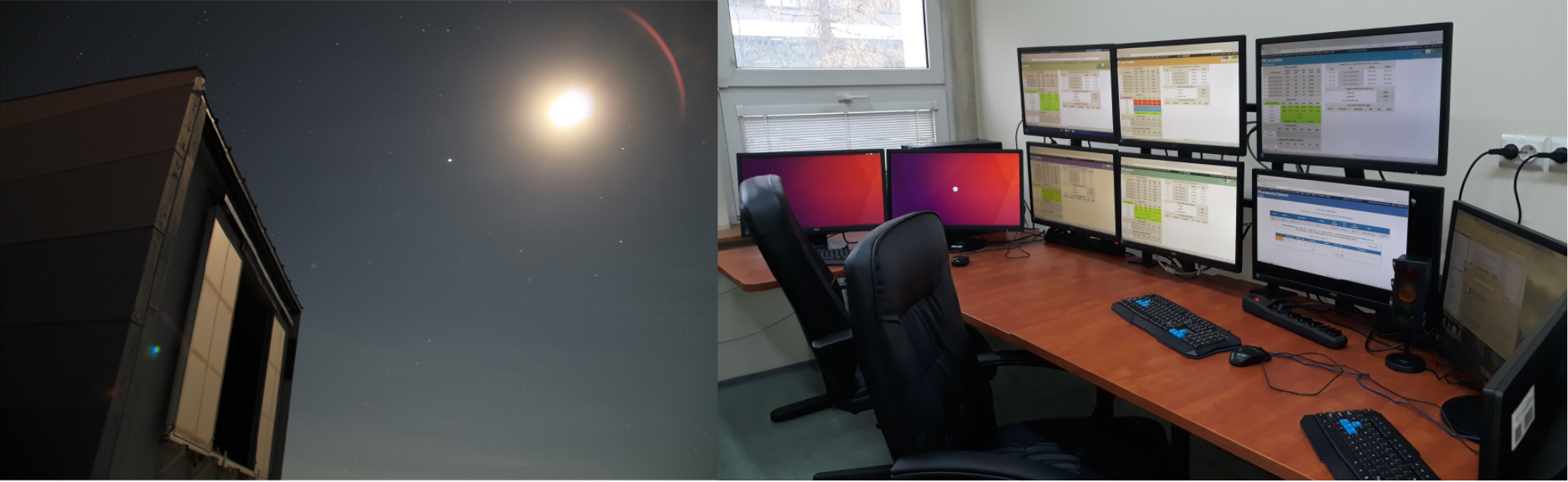}}
\caption{Left panel: Test of data acquisition with reduced PMTs gain performed at full moon.
Shown is the FD building with open shutters during the test. The full Moon and Jupiter are clearly visible on the sky above the building; Right panel: One of the FD remote-control rooms of the Auger Observatory.\label{FD-moon}}
\end{figure}

Out of the composition-sensitive observables, the atmospheric depth of maximum of shower development, X$_{\rm{max}}$, is currently the one with the smallest systematic uncertainties  
%observed directly by the FD) provides the smallest systematic uncertainties
and the most direct link to the mass distribution of the primary particles \cite{7a,7b}. 
It is best determined from the longitudinal profile of the shower, recorded using the fluorescence telescopes. 
Unfortunately,
the standard Fluorescence Detector operation is 
%\textcolor{red}{\sout{scheduled}}
performed only 
%\textcolor{red}{\sout{to}during}
during
nights with a low sky background, i.e. nights 
with 
%the Moon below the horizon for more than 3 hours \textcolor{red}{\sout{and}or} 
the illuminated fraction of the Moon below 70$\%$. The data acquisition is performed 
%only 
during astronomical nights (when the Sun is more than 18$^{\circ}$ below the horizon) and in good weather conditions. These all together inevitably limit the FD's duty cycle to $\sim$13$\%$. 
%Unfortunately, 
Due to the small duty cycle of the FD and a rapid decrease of the  cosmic ray flux with energy, the sufficient statistics of the recorded FD events,
%are available only below the energy of the flux suppression.
%Since the flux of the cosmic rays decreases rapidly with energy, statistically significant sample of high-quality hybrid data (i.e. showers seen by both the SD and FD), 
with direct optical observations of the shower maximum X$_{\rm{max}}$, 
%is currently the one with the smallest systematic uncertainties and the most direct link to the mass distribution of the primary particles \cite{7a,7b}. 
cover only energies up to 
%$\sim 19^{19.5}$ 
$\sim$40~EeV, i.e. just below the flux suppression region. 
%At the same time, the SD data extends to beyond 100 EeV. However, determining the composition of cosmic rays based solely on the SD (without the scintillator detectors, which are being installed currently) has been so far not very accurate due to large systematic uncertainties related to modeling hadronic showers and to limitations of reconstruction algorithms.

%It is best determined from the longitudinal profile of the shower, recorded using the fluorescence telescopes. Unfortunately, due to the small duty cycle of the FD, the sufficient statistics of the recorded FD events are available only below the energy of the flux suppression. Therefore, there is no easy way, at the moment, to settle the nature of the flux suppression. 
%A possible solution to this problem is to increase the measurement capability of SD, in particular by improving the ability to separate the lunar and electromagnetic components of the extensive showers

Hence the idea to extend the operation time of the fluorescence detector and to collect data even when the night sky background is high. The FD up-time can be increased if the restrictions on the illuminated fraction of the moon and its presence above the horizon are relaxed, and by including the astronomical twilight for the observation time (i.e. the time intervals when the Sun is between $12^{\circ}$
and $18^{\circ}$ below the horizon). No changes in the FD hardware are
%will be 
needed for this purpose.

%For FD to operate at night, when diffuse moonlight is increasing the background of the night sky, it is necessary to decrease the gain
%The conducted analyzes show that the most energetic events can be measured even with a 10-fold increase in the background
To operate the FD during nights when 
diffuse
%scattered 
moonlight increases 
%the night 
night sky background, it is necessary to decrease the gain of photomultipliers installed in the FD camera to avoid their damage. This in turn 
leads to degradation of the quality of collected data (especially for low energy events).
%reduces the efficiency of detection and reconstruction of extensive air showers. 
However, analyses performed indicate that the most energetic events 
%(i.e. cosmic rays with the energy above 10 EeV) 
can be measured even with the background increased by a factor of 10 (with appropriately reduced PMTs gain), while still keeping almost 100$\%$ selection efficiency as well as a good energy and X$_{\rm{max}}$  resolutions. This will allow to increase the FD duty cycle to $\sim$25$\%$ and thus
%at the same time
almost double the rate of collecting good-quality hybrid data above the energy of $\sim$40 EeV, providing (by increasing statistics) the valuable mass composition measurements in the flux suppression region. %These measurements can be later used as the cross-check of the other methods of mass composition determination. 
%The 
%same
%performed 
%analyses have also shown that 
%On the other hand, the efficiency of detection of showers with lower energies
%, during nights with 
%high 
%night 
%sky background, 
%will be heavily suppressed.
%However it is not important, since 
%their 
%the shower 
%statistics 
%for energies below the flux suppression region
%at lower energies 
%is already relatively large.
%However, this is not important as 
%their statistics are already relatively large.
%for lower energies we already have relatively large statistics.

The FD PMTs have been already extensively tested at the nominal and lowered gain levels in a dark chamber, 
showing a linear response in the considered gain range. Moreover, a dedicated test at the Pierre Auger Observatory was performed: in one of the telescopes the PMTs gain was reduced by a factor of 10, i.e. from the nominal gain of 50 000 to 5000, and showers with higher energies could be easily recorded
%seen 
%recorded at near full moon
%seen 
%even 
at full moon
%during the full moon 
%night
%night with high night sky background (near the full moon)
(see Fig.~\ref{FD-moon}, left panel). 
%More detailed simulations are still needed to establish the optimal gain of PMTs.

The extension of the FD up-time will result in an increase of the number and length of FD data acquisition shifts. A recently developed technique of remote shifts, which uses remote-control rooms at distant institutions that are connected to the Observatory via the Internet network, will significantly help in FD operation.
%by reducing the number of the on-site staff needed during the shift. 
%with performance operation 
%of the FD in the extended mode of the FD operation.
Currently,
there are several remote-control rooms being operational in the Auger Collaboration. One of such rooms is shown in Fig.~\ref{FD-moon} (right panel).

The extended mode of the FD operation will allow us to double the rate of accumulating high-quality hybrid
%X$_{\rm{max}}$ 
data (with directly measured X$_{\rm{max}}$) at the highest energies, providing the cross-check with the mass composition study performed by the upgraded Surface Detector.
%It will help us to study the mass composition of cosmic rays and possible elucidate the origin of the UHECRs flux suppression.
%(above $\sim$40 EeV).

\section{Performance of the AugerPrime Detector}

\begin{figure}[b]
\centerline{\includegraphics[width=7cm]{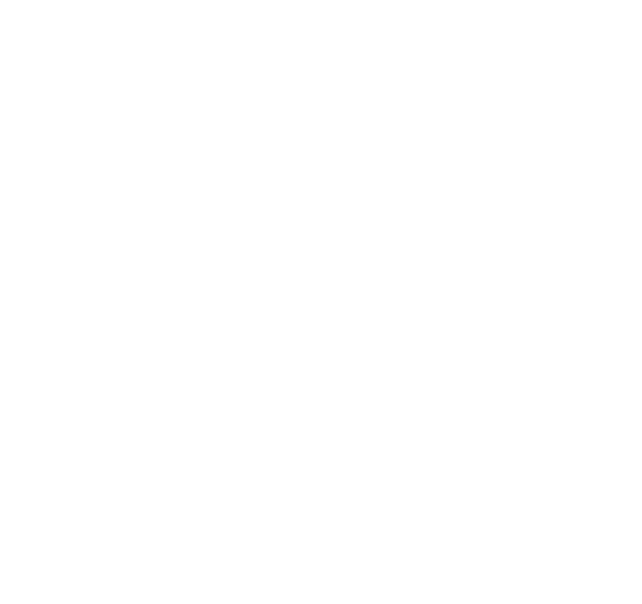}}
\caption{Significance of distinguishing between scenarios "1p" and "1" (i.e. between maximum rigidity scenario with
and without additional 10$\%$ fraction of protons, respectively) as a function of
the operation time of AugerPrime.
%Figures taken from Ref. \citen{}.
\label{10}}
\end{figure}

One of the key questions concerning the physics reach of the AugerPrime is its sensitivity for mass composition, in particular the feasibility to detect a fraction of protons as small as 10$\%$ at the highest energies. This is not easy to demonstrate, since we do not know what composition to expect. Nevertheless, one can select 
%For this reason, 
two exemplary 
benchmark
%simplified 
models of cosmic ray composition
%were selected 
for comparison to address this problem.
%For this reason two benchmark simplified models have been chosen to compare. 
One of them 
%models 
is a maximum rigidity scenario ("scenario 1"), corresponding to the best fit to the Auger cosmic ray flux and mass composition data 
%(atmospheric depth of shower maximum (X$_\rm{max}$) and its fluctuation) 
%above 
for $\rm{E}>10^{18.7}$ eV\cite{Combinedfit}.
%resulting from the numerical fitting  
%following the 
%method described in 
%(see Ref.~\citen{Combinedfit}). 
%The second scenario ("scenario 1p") is constructed from the first one by artificially adding a contribution of 10$\%$ protons to it.
The second scenario ("scenario 1p") is constructed on the basis of the first one by artificially adding a contribution of 10$\%$ protons to it. 
Two sets of mock data with a cosmic ray mass composition corresponding to these scenarios were generated and used to simulate the response of the upgraded Surface Detector (not including the radio detector). 
%Two sets of mock data, with the cosmic ray mass composition corresponding to the mentioned scenarios, have been generated and used to simulate response of the upgraded Surface Detector 
%(WCD and SSD). 
As a result, using the information from the SSD combined with the WCD data, the distributions of
%after applying the reconstruction algorithms with combined data provided by the WCD and SSD detectors,
depth of shower maximum (X$_{\rm{max}}$, measured indirectly), relative number of muons (R$_{\mu\rm{,38}}$),
%along with
and their fluctuations 
%of these observables 
(RMS(X$_{\rm{max}}$) and RMS(R$_{\mu\rm{,38}}$)) have been obtained
%simulated showers from 
for both sample datasets.
%both mock data sets. 
%For each scenario and observable, the mean of 
%along with the fluctuations of these observables, RMS(X$_{\rm{max}}$) and RMS(R$_{\mu\rm{,38}}$).
%have been also obtained.

\begin{figure}[h]
\centerline{\includegraphics[width=12.9cm]{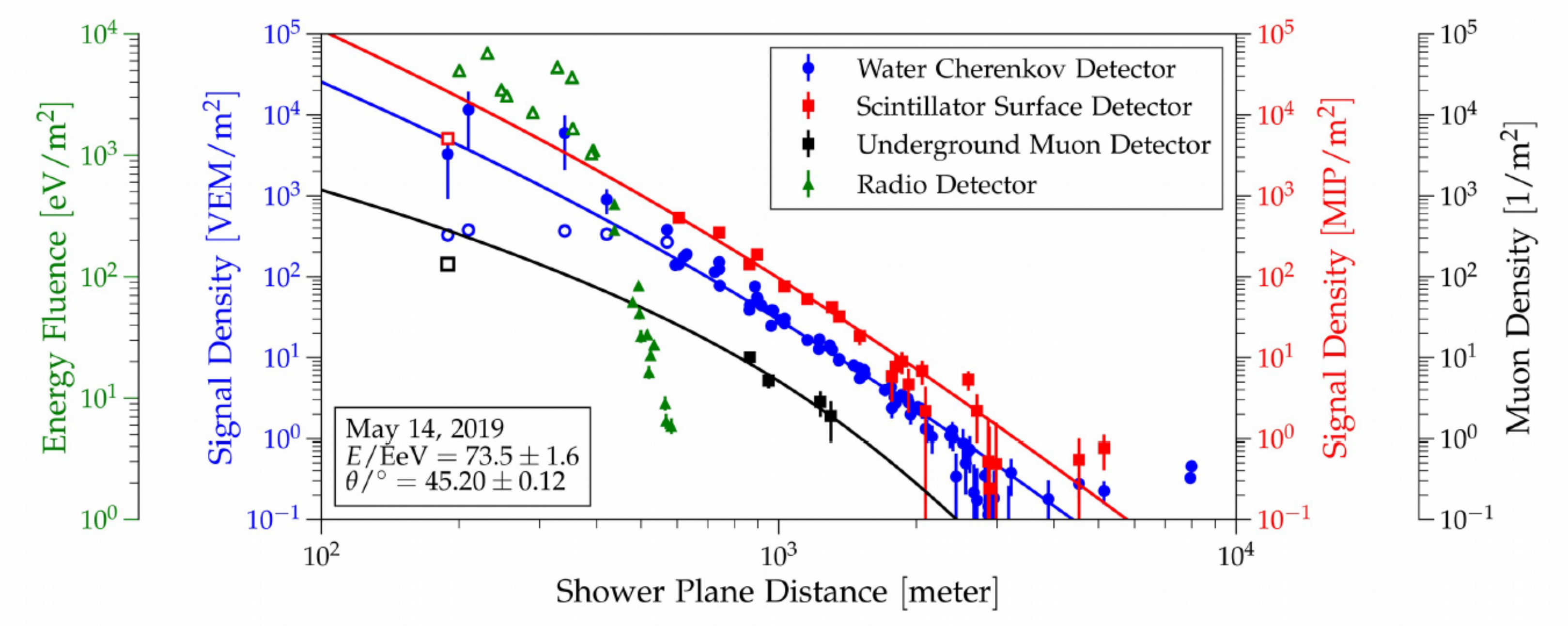}}
\caption{An example of a real multi-hybrid event, i.e. event observed simultaneously by various detectors of the Pierre Auger Observatory. Shown are lateral distributions of signals measured by WCD (blue), SSD (red), UMD (black), and RD (green), as a function of the distance to the shower core. \label{super-hybrid}}
\end{figure}

The difference $\sigma$ between the means of these distributions 
%of these observables 
as obtained from the two scenarios ("1" and "1p") divided by
the statistical uncertainty as a function of the operation time of the upgraded detector
%, for each observable, 
is shown in Fig.~\ref{10}.
As we can see, the two scenarios can be distinguished with high
significance and statistics.
The mass composition separation power can be enhanced by constructing a combined significance.
Using the combination of all of the available observables, 
%We conclude that using this combination 
we should be able to reach 10$\%$ sensitivity for a proton fraction at the highest energies with more than 5$\sigma$ confidence level after 5 years of the AugerPrime operation (see the black line in Fig.~\ref{10}).

\section{Conclusions}
\label{conclusion}

AugerPrime, the ongoing upgrade of the Pierre Auger Observatory, has  been  designed
to provide definitive answers to some of the pressing questions on UHECRs. It will enhance our ability to measure the mass composition of cosmic rays at energies above the flux suppression region and will allow for a superior separation of the muonic and electromagnetic components of air showers. As a part of the Upgrade, construction of the world-largest radio detector (covering an area of 3000 km$^2$) has already started. Moreover, scintillator detectors installed on top of the surface detectors across the  entire  Observatory  will enable a direct comparison of the Auger measurements with those of the surface detectors of the Telescope Array experiment.

The upgraded Observatory will be a multi-hybrid cosmic ray detector, that will allow simultaneous measurement of extensive air showers with 
%\mycorr{water-Cherenkov detectors, surface scintillator detectors, radio detectors, underground muon detectors and fluorescence detectors}
water-Cherenkov, surface scintillator, radio, underground muon, and fluorescence detectors. Its capabilities is illustrated in Fig.~\ref{super-hybrid}, where a real multi-hybrid event with the measured lateral signals of all the surface techniques is shown.

AugerPrime will help us to elucidate the origin of the UHECR flux suppression (to differentiate between the GZK or maximum rigidity scenario), to get information on primary mass composition on an event-by-event basis, to determine the proton flux contribution (as small as 10$\%$) at the highest energies, and last but not least, to 
reduce systematic uncertainties related to modeling hadronic showers.
%understand origin of the muon excess seen in the data.
%improve our knowledge about hadronic interactions at the highest energies.

%%%%%%%%%%%%%

%The SSD assembly is completed now, and their deployment on the field will soon be finished.

\section*{Acknowledgments}

The successful installation, commissioning, and operation of the Pierre Auger Observatory would not have been possible without the strong commitment and effort from the technical and administrative staff in Malargue, and the financial support from a number of funding agencies in the participating countries, listed at \url{https://www.auger.org/index.php/about-us/funding-agencies}. In particular we
want to acknowledge support in Poland from National Science Centre grants No.
2016/23/B/ST9/01635 and No. 2020/39/B/ST9/01398, and also from Ministry of Science and Higher Education grant No. DIR/WK/2018/11.

\bibliographystyle{ws-ijmpa}
\bibliography{ref}

\end{document}